\documentclass{article}
\usepackage[utf8]{inputenc}
\usepackage{amsmath}
\usepackage{amssymb}
\usepackage{amsthm}
\usepackage{fullpage}
\usepackage{graphicx}
\usepackage{xcolor}
\usepackage[nocompress]{cite}
\usepackage{url}

\newcommand{\LCP}{\textnormal{\texttt{lcp}}}
\newcommand{\LZss}{\textnormal{LZ77}}
\newcommand{\LZse}{\textnormal{LZ78}}
\newcommand{\firstnode}{v_{\mathrm{first}}}
\newcommand{\lastnode}{v_{\mathrm{last}}}

\newcommand{\problem}{string indexing with compressed pattern problem}
\newcommand{\substring}[3]{#1\left[#2,~#3\right]}
\newcommand{\LZ}[1]{\textnormal{LZ}\left(#1\right)}
\newcommand{\loc}{\texttt{locus}}
\newcommand{\Per}{\rho}
\newtheorem{theorem}{Theorem}
\newtheorem{lemma}{Lemma}[section]

\theoremstyle{definition}

\title{String Indexing with Compressed Patterns\footnote{An extended abstract appeared at the \emph{37th Symposium on Theoretical Aspects of Computer Science (STACS2020).} \cite{stacs}. The full version is published in {ACM} Trans. Algorithms (2023) \cite{DBLP:journals/talg/BilleGS23}.}}
\author{Philip Bille \\\texttt{phbi@dtu.dk} \and Inge Li G{\o}rtz \\\texttt{inge@dtu.dk} \and Teresa Anna Steiner \\\texttt{terst@dtu.dk}}

\date{}

\begin{document}

\maketitle

\begin{abstract}
Given a string $S$ of length $n$, the classic string indexing problem is to preprocess $S$ into a compact data structure that supports efficient subsequent pattern queries. In this paper we consider the basic variant where the pattern is given in compressed form and the goal is to achieve query time that is fast in terms of the compressed size of the pattern. This captures the common client-server scenario, where a client submits a query and communicates it in compressed form to a server. Instead of the server decompressing the query before processing it, we consider how to efficiently process the compressed query directly. Our main result is a novel linear space data structure that achieves near-optimal query time for patterns compressed with the classic Lempel-Ziv 1977 (LZ77) compression scheme. Along the way we develop several data structural techniques of independent interest, including a novel data structure that compactly encodes all \LZss\ compressed suffixes of a string in linear space and a general decomposition of tries that reduces the search time from logarithmic in the size of the trie to logarithmic in the length of the pattern. 
\end{abstract}
\section{Introduction}

The string indexing problem is to preprocess a string~$S$ into a compact data structure that supports efficient subsequent pattern matching queries, that is, given a pattern string $P$, report all occurrences of $P$ within~$S$. In this paper, we introduce a basic  variant of string indexing, called the \emph{\problem}, where the pattern $P$ is given in compressed form and we want to answer the query without decompressing $P$. The goal is to obtain a compact structure while achieving fast query times in terms of the compressed size of $P$.

The \problem\ captures the following common client-server scenario: a client submits a query and sends it to a server which processes the query. To minimize communication time and bandwidth the query is sent in compressed form. Naively, the server will then have to decompress the query and then process it. With an efficient solution to the \problem\ we can eliminate the overhead decompression and speed up queries by exploiting repetitions in pattern strings.

We focus on the classic Lempel-Ziv 1977 (\LZss)~\cite{ziv1977universal} compression scheme. Note that since the size of an \LZss\ compressed string is a lower bound for many other compression schemes (such as all grammar-based compression schemes) our results can be adapted to such compression schemes by recompressing the pattern string. To state the bounds, let $n$ be the length of $S$, $m$ be the length of $P$, and $z$ be the \LZss\ compressed length of $P$. Naively, we can solve the \problem\ by using a suffix tree of $S$ as our data structure and answering queries by first decompressing them and then traversing the suffix tree with the uncompressed pattern. This leads to a solution with $O(n)$ space and $O(m + \mathrm{occ})$ query time. At the other extreme, we can store a trie of all the \LZss\ compressed suffixes of $S$ together with a simple tabulation, leading to a solution with  $O(n^3)$ space and $O(z + \mathrm{occ})$ query time (see discussion in Section~\ref{sec:phrase_trie}). 

While the opposite problem, where the indexed string $S$ is compressed and the pattern $P$ is uncompressed, is well-studied~\cite{karkkainen1996lempel,karkkainen1998lempel,ferragina2000opportunistic,makinen2000compact,grossi2005compressed,ferragina2001experimental,ferragina2005indexing,grossi2003high,grossi2004indexing,ferragina2007compressed,navarro2007compressed,makinen2010storage,claude2012improved,kreft2013compressing,maruyama2013100,gagie2014lz77,belazzougui2014alphabet,bille2018time} (see also the surveys~\cite{navarro2007compressed,navarro2012indexing,navarro2016compact,gagie2015searching}), little is known about  the \problem. 
As an intermediate result in their paper on indexed multi-pattern matching,  Gagie~et~al.~\cite{GagieKKMST12} give a  data structure using $nH_k(S)+o(n(H_k(S)+1))$ bits, which can find the suffix array interval for an \LZss-compressed pattern in  $O(z\log^2m\log^{1+\epsilon}n)$ time, where $z$ is the number of phrases in the \LZss\ compression of the pattern. Their strategy is to convert the \LZss\ compression to a straight-line program (SLP) and use iterative merging of suffix array intervals for concatenated strings. Combined with the more recent data structure by Fischer et al. \cite{FischerKK16}, this implies a solution to the \problem\ using linear space and $O(z\log (m/z) \log \log n + \mathrm{occ})$ query time. However, since these solutions convert the \LZss\ compression to an SLP, the size of the SLP compression is a bottleneck for the query time. The best-known conversion achieves an SLP of size $O(z\log (m/z))$ and the size of the smallest SLP is lower bounded by $\Omega(z\log m/\log\log m)$~\cite{charikar05}.

We present a new solution to the \problem\ achieving the following bound: 
\begin{theorem}
\label{thm:mainresult}
We can solve the \problem\ for \LZss-compressed patterns in~$O(n)$ space and~$O(z+\log m + \mathrm{occ})$ time, where $n$ is the length of the indexing string, $m$ is the length of the pattern, and $z$ is the number of phrases in the \LZss\ compressed pattern.
\end{theorem}
Since any solution must use at least $\Omega(z + \mathrm{occ})$ time to read the input and report the occurrences, the time bound in Theorem~\ref{thm:mainresult} is optimal within an additive $O(\log m)$ term. In the common case when $z = \Omega(\log m)$ or if we consider \LZss\ without \emph{self-references} the time bound is optimal. For simplicity, we focus on reporting queries, but the result is straightforward to extend to also support \emph{existential queries} (decide if the pattern occurs in $S$) and \emph{counting queries} (count the number of occurrences of the pattern in $S$) in $O(z+\log m)$ time and the same space.

To achieve Theorem~\ref{thm:mainresult} we develop several data structural techniques of independent interest. These include a compact data structure that encodes all \LZss\ compressed suffixes of a string in linear space in the length of the string and a general decomposition of tries that reduces the search time from logarithmic in the size of the trie to logarithmic in the length of the pattern. 

Let $z_i$ be the number of phrases in the LZ77 compression of the $i$th suffix of $S$, for all $0\leq i\leq n-1$. We  show how to build the data structure from Theorem \ref{thm:mainresult} in $O(n\log n+\sum_{i=0}^{n-1}z_i\log \log n)$ expected time for \LZss\ without self-references, and $O(n^2+\sum_{i=0}^{n-1}z_i\log \log n)$ expected time if we allow self-referencing. Further, we show how to extend our results to the Lempel-Ziv 1978 (LZ78)~\cite{ZivL78} compression scheme in the same complexities.

A related problem has been studied in a line of work on \emph{fully compressed pattern matching}, where the goal is to locate a pattern $P$ within a string $S$ when both are given in compressed form~\cite{GasieniecR99,HiraoSTA00,InenagaST05,Gawrychowski12,Jez15}.

The paper is organized as follows. In Section~\ref{sec:preliminaries} we recall basic string data structures and \LZss\ compression. In Section~\ref{sec:phrase_trie} we present a simple $O(n^2)$ space and $O(z + \log n + \mathrm{occ})$ time data structure that forms the basis of our solutions in the following sections. In Section~\ref{sec:compact_phrase_trie} we show how to achieve linear space with the same time complexity. In Sections~\ref{sec:slice_tree} and \ref{sec:saving_space} we show how to improve the $\log n$ term to $\log m$, proving the main theorem. In Section~\ref{sec:lz78}, we show how to extend these results to the \LZse\ compression scheme. 

\section{Preliminaries}\label{sec:preliminaries}
A string~$S$ of length~$n$ is a sequence $S[0]\cdots S[n-1]$ of $n$ characters drawn from an alphabet $\Sigma$. The string $S[i]\cdots S[j-1]$ 
denoted $\substring{S}{i}{j}$ is called a \emph{substring} of $S$. The substrings $\substring{S}{0}{j}$ and $\substring{S}{i}{n}$  are called the $j^{th}$ \emph{prefix} and  $i^{th}$ \emph{suffix} of $S$, respectively. We will sometimes use $S_i$ to denote the $i^{th}$ suffix of $S$.
 
\paragraph{Longest Common Prefix}
    For two strings $S$ and~$S'$, the \emph{longest common prefix of }$S$ and~$S'$, denoted \LCP$(S,S')$, is the maximum $j\in \{0,\dots, \min\left(|S|, |S'|\right)\}$ such that $\substring{S}{0}{j}=\substring{S'}{0}{j}$.

Given a string $S$ of length $n$, there is a data structure of size $O(n)$ that answers \LCP-queries for any two suffixes of $S$ in constant time by storing a suffix tree combined with an efficient nearest common ancestor (NCA) data
structure~\cite{harel1984fast, weiner1973linear}.

\paragraph{Compact Tries}
Let $D$ be a set of strings $S^1, \ldots, S^l$, and assume without loss of generality that the strings in $D$ are prefix free (if they are not, append each string with a special character \$ which is not in the alphabet).
A \emph{compact trie} for $D$ is a rooted labeled tree $T_D$, with the following properties: The label on each edge is a substring of one or more $S^i$. Each root-to-leaf path represents a string in the set (obtained by concatenating the labels on the edges of the path), and for every string there is a leaf corresponding to that string. Common prefixes of two strings share the same path maximally, and all internal vertices have at least two children. 

The compact trie has $O(l)$ nodes and edges and a total space complexity of $O\left(\sum_{i=1}^l |S^i|\right)$. The position in the trie that corresponds to the maximum longest common prefix of a pattern $P$ of length $m$ and any $S^i$ can be found in $O(m)$ time. For a position $p$ in the tree, which can be either a node or a position within the label of an edge, let str($p$) denote the string obtained by concatenating the labels on the path from the root to $p$. The locus of a string $P$ in $T_D$, denoted $\loc(P)$, is the deepest position $p$ in the tree such that str($p$) is a prefix of $P$. 
A compact trie on the suffixes of a string $S$ is called the \emph{suffix tree} of $S$ and can be stored in linear space~\cite{weiner1973linear}.  The \emph{suffix array} stores the starting positions of the suffixes in the string in lexicographic order. If at every node in the suffix tree its children are stored in lexicographic order, the order of the suffix array corresponds to the order of the leaves in the suffix tree.

\paragraph{LZ77}\label{subsec:lz77}
Given an input string $S$ of length $n$, the \LZss\ parsing divides $S$ into $z$ substrings $f_1, f_2,\ldots,f_z$, called phrases, in a greedy left-to-right order. The $i^{th}$ phrase $f_i$, starting at position $p_i$ is either (a) the first occurrence of a character in $S$ or (b) the longest substring that has at least one occurrence starting to the left of $p_i$.
If there is more than one occurrence, we assume that the choice is made in a consistent way.
To compress $S$, we can then replace each phrase $f_i$ of type (b) with a pair $(r_i,l_i)$ such that $r_i$ is the distance from $p_i$ to the start of the previous occurrence, and $l_i$ is the length of the phrase. If $l_i> r_i$, we call $f_i$ \emph{self-referencing}. The occurrence of $f_i$ at position $p_i-r_i$ is called the \emph{source} of the phrase. (This is actually the \LZss-variant of Storer and Szymanski~\cite{storer1982data}; the original one~\cite{ziv1977universal} adds a character to each phrase so that it outputs triples instead of tuples.) We have $z = O(n/\log_\sigma n)$. Furthermore, 
if self-references are not  allowed then $z = \Omega(\log n)$, whereas $z = \Omega(1)$ for self-referential parses. 

Every \LZss-compressed string is a string over the extended alphabet which consists of all possible \LZss\ phrases. For any string $T$ we denote this string by $\LZ{T}$.

\section{A Simple Data Structure}\label{sec:phrase_trie}
In this section, we will define a data structure that allows us to solve the \problem\ in $O(n^2)$ space and $O(z+\log n + \mathrm{occ})$ time, or $O(n^3)$ space and $O(z+\mathrm{occ})$ time. This data structure forms the basis of our solution.

\paragraph{The Phrase Trie}

\begin{figure}[t]
    \centering
    \includegraphics[width=0.8\textwidth]{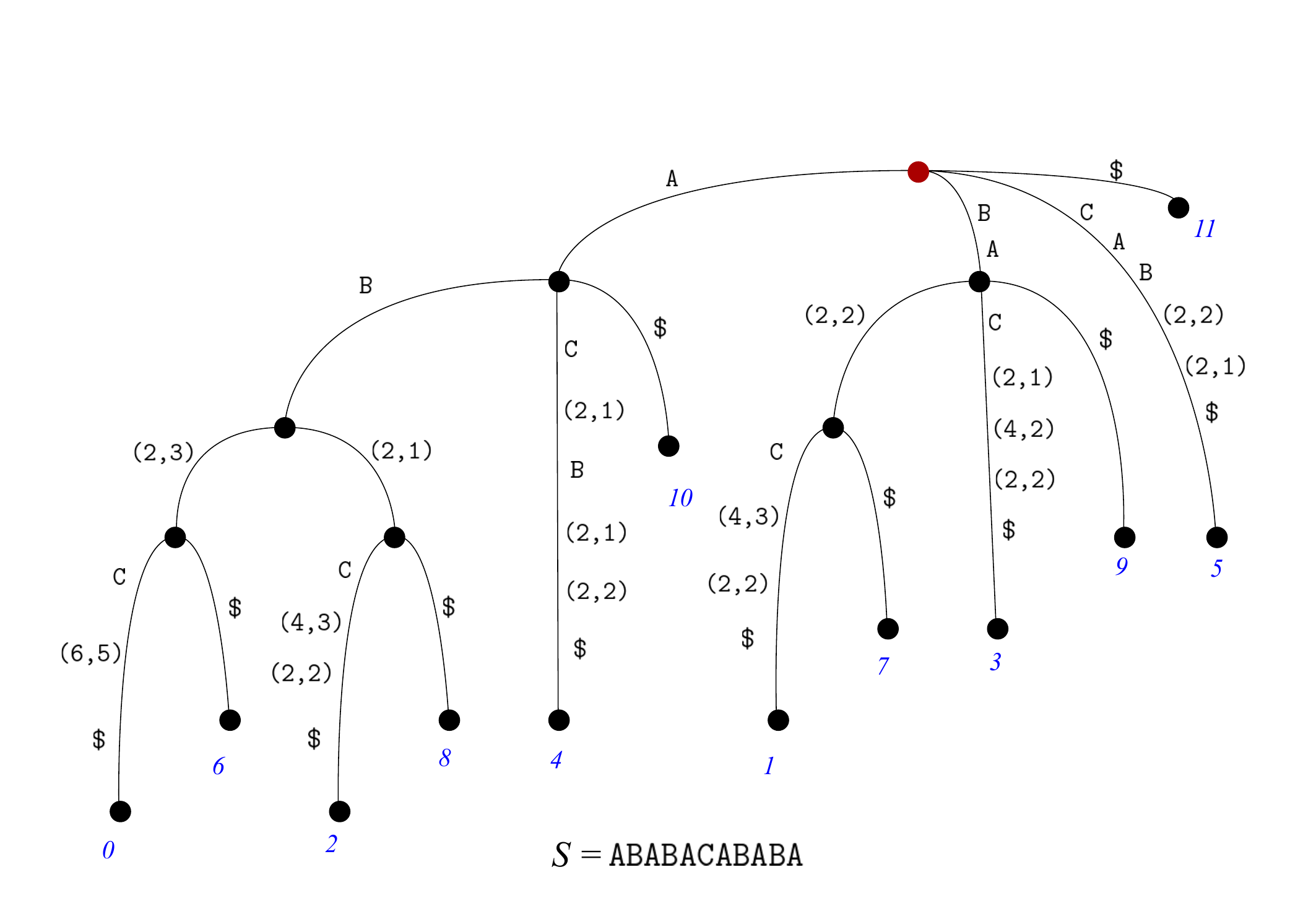}
    \caption{The~phrase trie for the string \texttt{ABABACABABA\$}. In this example, the leaves are sorted according to the lexicographic order of the originial suffixes. For instance the $6^{th}$ suffix \texttt{ABABA\$} has the \LZss\ parse~\texttt{A~B~(2,3)~\$}, and this string corresponds to the concatenation of labels on the path from the root to the second leaf.}
    \label{fig:phrase_trie}
\end{figure}

The \emph{phrase trie of a string} $S$ is defined as the compact trie over the set of strings $\{\LZ{S_i\$},i=0,\dots,|S|-1\}\cup\{\$\}$, that is, the \LZss\ parses of all suffixes of $S$ appended by a new symbol $\$ $ which is lexicographically greater than any letter in the alphabet. For an example see Figure \ref{fig:phrase_trie}.

The phrase trie for a string $S$ of length $n$ has $n+1$ leaves, one corresponding to every suffix of $S\$ $. Similarly as in the suffix tree, every internal node defines a consecutive range within the suffix array. Since every node has at least 2 children the number of nodes and edges is $O(n)$. However, we have to store labels corresponding to the \LZss\ parses of the suffixes of $S$, using worst case $\Theta(n^2)$ space.

\LZss\ has the property that for two strings  whose prefixes  match up to some position $\ell$, the \LZss-compression of the two strings will be the same up to (not necessarily including) the phrase that contains position $\ell$. 
As such, we can use the phrase trie to find the suffix $S_i$ of $S$ for which the \LZss-compression of the pattern $P$ agrees with the \LZss-compression of $S_i$ for as long as possible. Assuming they match for $k-1$ phrases, the longest match of $P$ in $S$ ends within the $k^{th}$ phrase. 
Now, the problem reduces to the following: given the set of suffixes for which the \LZss-compression maximally agrees with $P$, assuming they match for $k-1$ phrases and until position $p$, find the subset of those suffixes for which the $k^{th}$ phrase agrees longest with the $k^{th}$ phrase of $P$. For a fixed string $S$ of length $n$, there are at most $n$ different choices for position $p$, and at most $n^2$ different choices for the encoding of the next phrase in the pattern. As such, we can store the pre-computed solutions for all cases in a table using an additional $O(n^3)$ space for solving the problem in $O(z+\mathrm{occ})$ time.
Instead, we will store a linear space and constant time \LCP\ data structure for suffixes of $S$ and show that given the first phrase where the suffix $S_i$ and the string $P$ mismatch, we can find the \LCP\ of $P$ and $S_i$ by finding the \LCP\ of two suffixes of $S$. 
This will allow us to search for the longest match of $P$ in $S$ in at most $O(\log n)$ extra time.

\paragraph{Longest Common Prefixes in LZ77-Compressed Strings}

\begin{figure}
    \centering
    \includegraphics[width=0.5\textwidth]{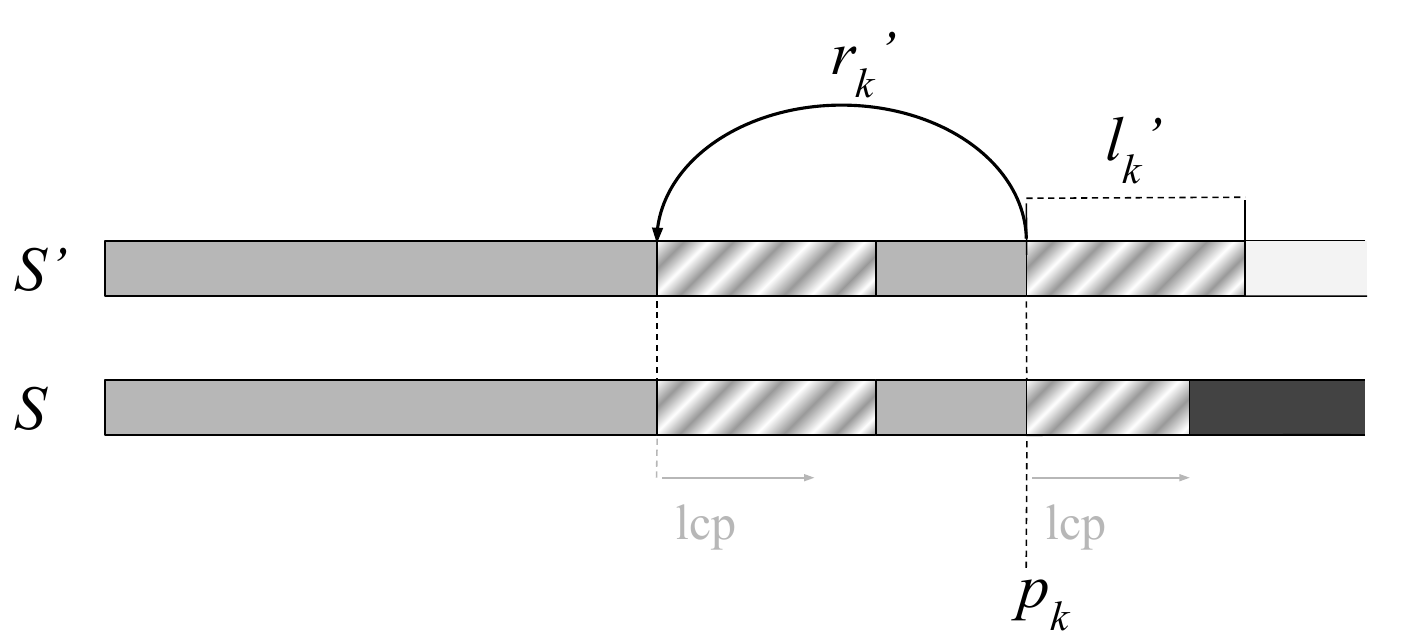}
    \caption{The~$k^{th}$ phrase in~$S'$ is copied from position~$p_k-r'_k$, at which point~$S$ and~$S'$ are identical; the \LCP\ value gives how far~$p_k$ and~$p_k-r'_k$ match in~$S$.}
    \label{fig:lcemma}
\end{figure}

We will use an intuitive property about \LZss-compressed strings: assuming two strings match up until a certain phrase~$k-1$, we can reduce the task of finding the~$\LCP$ of the two strings to the task of finding the longest common prefix between two suffixes of one of the strings. 
This property is summarized in the following lemma (see also Figure~\ref{fig:lcemma}):

\begin{lemma}\label{lcemma}
        Let 
           ~$S=f_1~f_2~\cdots~f_z$ and
           ~$S'=f'_1~f'_2~\cdots~f'_{z'}$ be two strings parsed into \LZss\ phrases, where~$f_1=f'_1$,~$f_2=f'_2,\dots, f_{k-1}=f'_{k-1}$ for some $k$. Let $p_k$ be the starting position of $f_k$ and $f'_k$. If~$f'_k$ is a  phrase represented by a pair $(r'_k, l'_k)$ the following holds:
            \begin{align}\label{form:lcp}\LCP\left(S,S'\right)\geq p_k+\min\left(\LCP\left(\substring{S}{p_k}{n},\substring{S}{p_k-r'_k}{n}\right),l'_k\right).\end{align}
            Furthermore, if~$f_k\neq f'_k$, equality holds in (\ref{form:lcp}).
    \end{lemma}

    \begin{proof}

    To prove the lower bound stated in (\ref{form:lcp}), we will show by induction that for any $i\leq \min\left(\LCP\left(\substring{S}{p_k}{n},\substring{S}{p_k-r'_k}{n}\right),l'_k\right)$, we have that~$S[p_k+i-1]=S'[p_k+i-1]$. For~$i=0$ this is true since~$S$ and~$S'$ are the same up until position~$p_k-1$. For the induction step assume it is true for all~$i_0<i$. We then have 
    \begin{align}
        S'[p_k+i-1] &= S'[p_k-r'_k+i-1]\label{lcemma11}\\
        &=S[p_k-r'_k+i-1]\label{lcemma12}\\
        &=S[p_k+i-1]\label{lcemma13},
    \end{align}
    where (\ref{lcemma11}) follows from~$i\leq l'_k$ and because $p_k-r'_k$ is the source of phrase $f'_k$, (\ref{lcemma12}) follows from the induction hypothesis, and (\ref{lcemma13}) follows from~$i\leq \LCP\left(\substring{S}{p_k}{n},\substring{S}{p_k-r'_k}{n}\right)$.

    To show equality in the case where~$f_k\neq f'_k$, let~$t=\min\left(\LCP\left(\substring{S}{p_k}{n},\substring{S}{p_k-r'_k}{n}\right),l'_k\right)$. We will show that~$S[p_k+t]\neq S'[p_k+t]$. There are two cases:

        For~$t = \LCP\left(\substring{S}{p_k}{n},\substring{S}{p_k-r'_k}{n}\right)< l'_k$, note that $S[p_k-r'_k+t]\neq S[p_k+t]$. From (\ref{form:lcp}) we know that $S'[p_k-r'_k+t]=S[p_k-r'_k+t]$, and therefore we have  $ S'[p_k+t]=S'[p_k-r'_k+t]=S[p_k-r'_k+t]\neq S[p_k+t]$.

        For~$l'_k\leq\LCP\left(\substring{S}{p_k}{n},\substring{S}{p_k-r'_k}{n}\right)$, note that by (\ref{form:lcp}), we know that~$S$ and~$S'$ have an \LCP~of length at least~$p_k+t$. If~$t\geq l_k$, then by the uniqueness of the greedy left-to-right parsing, the $k^{th}$ phrase of~$S$ and~$S'$ would be the same, contradicting our condition. Otherwise, we have~$l_k>t = l'_k$. This together with~(\ref{form:lcp}) implies $S[p_k+i]=S[p_k+i-r_k]=S'[p_k+i-r_k]$ for every $i=0,\dots,t$, since $r_k\geq 1$. By the greedy parsing property and since $l'_k=t$ we know that $S'[p_k+t-r_k]\neq S'[p_k+t]$ and so $S[p_k+t]\neq S'[p_k+t]$. 
    \end{proof}

\subsection{The Data Structure}
Additionally to storing the phrase trie of $S$, we store the suffix array of~$S$, and for every node in the phrase trie, the range of the leaves below it in the suffix array. Finally, we store a linear space and constant time data structure for answering \LCP-queries for suffixes of $S$.

\subsection{Algorithm}
The algorithm we describe in this section, as well as all solutions presented later in the paper, actually solve the more general problem of finding all occurrences of the longest prefix of $P$ that is a substring of $S$. 
We begin by matching $\LZ{P}$ as far as possible in the phrase trie. Let $v = \loc(\LZ{P})$. 
Let $k$ be the first phrase in $\LZ{P}$ that does not match any of the next phrases in the trie. If $v$ is a node then set $w=v$, otherwise let $w$ be the first node below $v$. We proceed as follows:
\begin{itemize}
        \item If the~$k^{th}$ phrase in~$P$ is a single letter, we return~$p_k$ as the length of the match and the interval of positions stored at $w$. 
        \item If the~$k^{th}$ phrase is represented by $(r_k,l_k)$ then there are two cases:
        \begin{itemize}
        \item If~$v$ is on an edge, let~$S_i$ be the suffix corresponding to any leaf below~$v$.
        We return $p_k+\min(\LCP\left(\substring{S}{i+p_k}{n},\substring{S}{i+p_k-r_k}{n}\right),l_k)$
        as the length of the match and the interval of positions stored at~$w$.
        \item If~$v$ is on a node, we do a binary search for the longest match in the range in the suffix array below~$v$, in the following way. For the suffix $S_i$ that corresponds to the index in the middle of the given range, we compute $\LCP(\substring{S}{i+p_k}{n},\substring{S}{i+p_k-r_k}{n})$. If this is greater than $l_k$ we stop the binary search.  Otherwise, we check if the next position in~suffix $S_i$ is lexicographically smaller or bigger than the next position in~$P$ to see whether we go left or right in the binary search. That is, let $t=p_k+\LCP(\substring{S}{i+p_k}{n},\substring{S}{i+p_k-r_k}{n})+1$.
        We compare $S_i[t]=S[i+t]$ with $P[t]=S_i[t-r_k]=S[i+t-r_k]$. If $P[t]$ is lexicographically smaller, we recurse on the left half of the current range, otherwise on the right. Throughout the process, we keep track of the longest match we have seen so far, since updating the search does not necessarily mean that a longer match can be found in the new interval. At the end of the search, we go to the longest match seen and check left and right for all occurrences, since there can be some that the binary search skipped: Multiple consecutive leaves can share the same longest prefix with $P$ while some of them are lexicographically smaller and some are lexicographically bigger.

    \end{itemize}
    \end{itemize}
    
\subsection{Correctness} 
The compact trie gives us the longest matching prefix of $\LZ{P}=f_1\dots f_{z_p}$ in the phrase trie. That is, we find all suffixes $S_i=f'_1 \cdots f'_{z_i}$ for $i=0,\dots,n-1$ such that $f_1=f'_1, \dots, f_{k-1}=f'_{k-1}$ and $f_k\neq f'_{k}$, and $k$ is maximal. By the uniqueness of parsing, the longest prefix of $P$ found in $S$ is the prefix of at least one of these suffixes.

Note that by the greedy parsing, the longest match of the $k^{th}$  phrase has to end before the next node in the trie. We argue the different cases:

    If the~$k^{th}$ phrase in~$P$ is a letter, it did not appear in~$P$ before. Thus, it never appeared in any of the suffixes we matched so far. Since the next phrase in the phrase trie is different, it is either a copied position, or a different letter. In any case, the next letter of any candidate suffix does not match the next letter in~$P$.  

    If $f_k$ is represented by $(r_k,l_k)$ there are two subcases.
    If~$v$ is on an edge, recall that~$S_i$ is the suffix corresponding to any leaf below the current position $v$. By Lemma~\ref{lcemma} and since~$S_i[p]=S[p+i]$ for any~$p$, we have that
    \begin{align*}\LCP\left(S_i,P\right)&=p_k+\min(\LCP\left(\substring{S_i}{p_k}{n},\substring{S_i}{p_k-r_k}{n}\right),l_k) \\
    &=p_k+\min(\LCP\left(\substring{S}{i+p_k}{n},\substring{S}{i+p_k-r_k}{n}\right),l_k).
    \end{align*}
As such, we return the correct length, and since the match ends on this edge the occurrences of the longest prefix of $P$ correspond to the suffix array interval stored at the next node below.

    If~$v$ is on a node we have, by the same argument as before,~$$\LCP\left(S_i,P\right)=p_k+\min(\LCP\left(\substring{S}{i+p_k}{n},\substring{S}{i+p_k-r_k}{n}\right),l_k),$$ for every~suffix $S_i$. 
  Further, because of the lexicographic order of the suffix array, we can binary search to find the leaf with the longest match: At any point in the search, when we compare with a suffix $S_i$ and if there exists another suffix that has a longer common prefix with $P$, it will be lexicographically smaller than suffix $S_i$ exactly if $P$ is lexicographically smaller than $S_i$. If we compare with a suffix $S_i$ that maximally shares a prefix with $P$, there might be other suffixes both lexicographically bigger and smaller than $S_i$ that share the same prefix. However, they will all be adjacent, and by checking the adjacent positions of the longest match in the suffix array we make sure to find all occurrences.

\subsection{Analysis}
The suffix array and the~\LCP\ data structure both use linear space in the size of~$S$. For the phrase trie, we store the \LZss-compressed suffixes of $S$, which use $O(\sum_{i=0}^{n-1} z_i)=O(n^2)$ space, where $z_i$ is the number of phrases used to compress suffix $S_i$.

For the time complexity, we use~$O(k)=O(z)$ time for matching the phrases in the trie.
In the worst case, that is, when the locus~$v$ is on a node, we need~$O(\log(\#\texttt{leaves~below } v))=O(\log n)$ 
constant time $\LCP$
queries. In total, we have a time complexity of~$O(z+\log n+\mathrm{occ})$.
In summary, we proved the following lemma.
\begin{lemma}
    The phrase trie solves the \problem\ in~$O(n^2)$ space and~$O(z+\log n + \mathrm{occ})$ time.
\end{lemma}

\subsection{Preprocessing} The suffix tree and suffix array can be constructed in time $O(\mathrm{sort}(n,\sigma))$, where $\sigma$ is the size of the alphabet and $\mathrm{sort}(n,\sigma)$ is the sorting complexity of sorting $n$ numbers from a universe of size $\sigma$. This is linear in $n$ for linear-sized alphabets \cite{weiner1973linear,DBLP:journals/jacm/KarkkainenSB06,journals/jacm/Farach-ColtonFM00}. To enable constant time access of the correct outgoing edge at each node, we use perfect hashing \cite{FKS1984}, which requires an additional $O(n)$ expected preprocessing time.
 The NCA data structure used for the \LCP\ data structure can be constructed in time $O(n)$ \cite{harel1984fast}.
 
 To construct the phrase trie, we need to find the \LZss\ parses of all the suffixes. To compute the \LZss\ compression of each suffix, we will use results by Keller et al.~\cite{DBLP:journals/tcs/KellerKFL14} for generalized substring compression. The data structure in \cite{DBLP:journals/tcs/KellerKFL14} uses a suffix tree augmented with constant amount of information per node, which can be constructed in linear time, plus an NCA data structure and a range successor data structure, to compress any substring $S[i,j]$ of $S$ in time $O(z_{i,j}\cdot Q_{\mathrm{succ}})$. Here, $z_{i,j}$ is the number of phrases in $S[i,j]$ and $Q_{\mathrm{succ}}$ is the query time for range successor. An $O(n\log \log n)$ space and $O(\log \log n)$ time range successor data structure can be built in $O(n\sqrt{\log n})$ time \cite{DBLP:conf/esa/Gao0N20}. Thus we can find the \LZss\ parses of all suffixes in time $O(n\sqrt{\log n}+ \sum_{i=0}^{n-1} z_i\log \log n)$. 
 Since we already constructed the suffix tree, we can assume we have the \LZss\ parses of all suffixes sorted by lexicographic order. Using perfect hashing again, we can build the phrase trie from those in $O(n+\sum_{i=0}^{n-1}z_i)$ expected time.

 Summing up, we can build the data structure in $O(n\sqrt{\log n}+\sum_{i=0}^{n-1} z_i\log \log n)$ expected time.

\section{Space Efficient Phrase Trie}\label{sec:compact_phrase_trie}
In this section, we show how to achieve the same functionality as the phrase trie while using linear space. The main idea is to store only one phrase per edge, and use Lemma~\ref{lcemma} to navigate along an edge. That is, we no longer store the entire \LZss-compressed suffixes of $S$.

\subsection{The Data Structure} 
\begin{figure}
    \centering
    \includegraphics[width=0.8\textwidth]{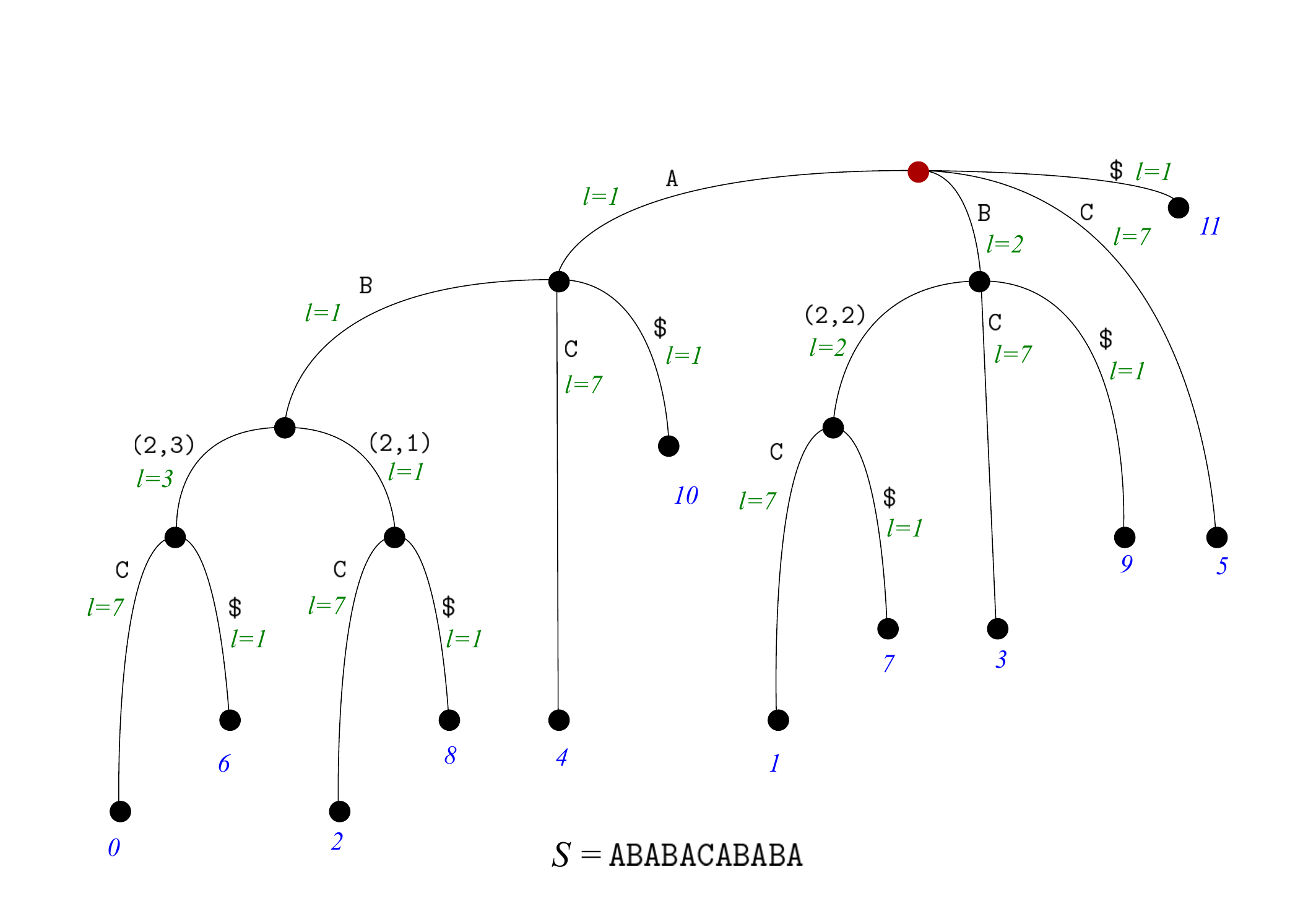}
    \caption{The~phrase trie for the string \texttt{ABABACABABA} using linear space.}
    \label{fig:compact_phrase_trie}
\end{figure}

We store a compact form of the phrase trie, which is essentially a blind trie version of the phrase trie. 
In contrast to the usual blind trie, we do not store the actual strings or the compressed strings on the side to verify, but show that the structure of the \LZss-compression scheme is enough to ensure navigating within the blind trie without false positives.
We store the following: We keep the tree structure of the phrase trie, and at each node, we keep a hash table, using perfect hashing~\cite{FKS1984}, where the keys are the first \LZss\ phrase of each outgoing edge. For each edge we store as additional information the length of the (uncompressed) substring on that edge and an arbitrarily chosen leaf below it. For an example see Figure \ref{fig:compact_phrase_trie}. As before, we additionally store the suffix array, the range within the suffix array for each node, and a linear-sized \LCP\ data structure for suffixes $S$.

\subsection{Algorithm}
 The algorithm proceeds as follows. We start the search at the root. Assume we have matched~$k-1$ phrases of~$P$ and the current position in the trie is a node $v$. To match the next phrase we check if the $k^{th}$ phrase in~$P$ is in the hash table of $v$.
    \begin{enumerate}
        \item If~it is not, we proceed exactly as in the previous section in the case where the locus is at a node. 
        \item If the $k^{th}$ phrase is present, let $e$ be the corresponding edge and let $i$ be the starting index of the leaf stored for $e$.
         Set $k=k+1$. We do the following until we reach the end of edge $e$ or get a mismatch. We differentiate between two cases.\label{edgemoving}
        \begin{itemize}
         \item The $k^{th}$ phrase in $P$ is a single letter $\alpha$:
        \begin{itemize}
            \item If $\alpha=S[i+p_k]$, we set $k=k+1$ and continue with the next phrase.
            \item If $\alpha\neq S[i+p_k]$, we stop and return $p_k$ as the length of the match together with the interval of occurrences stored at the next node below.
        \end{itemize}
        \item The $k^{th}$ phrase in $P$ is represented by $(r_k,l_k)$:
        \begin{itemize}
        \item If $\min(\LCP(\substring{S}{i+p_k}{n},\substring{S}{i+p_k-r_k}{n}),l_k)= l_k$, we set~$k=k+1$ and continue with the next phrase. 
        \item Otherwise, 
        we return~$p_k + \LCP(\substring{S}{i+p_k}{n},\substring{S}{i+p_k-r_k}{n})$ as the length of the match, with the interval of positions stored at the next node.
        \end{itemize}
        \end{itemize}
        If we reach the end of an edge,
        we go to the next node below and continue in the same way.
    \end{enumerate}

\paragraph{Correctness}
The correctness follows from the previous section together with Lemma~\ref{lcemma}, since we always keep the invariant that when we process the $k^{th}$ phrase, we already matched the $k-1$ previous ones.

\paragraph{Analysis}
The space complexity is linear since the compact phrase trie has~$O(n)$ nodes and edges and stores constant information per node and edge, using perfect hashing.

The time complexity is the same as in the previous section, since for matching full phrases, we use at most one constant time lookup in the hash table and one constant time \LCP~query per phrase in~$P$. As before, the worst case for matching the~$k^{th}$ phrase is having to do a binary search, using~$O(\log n)$ constant time \LCP~queries. In summary, this gives the following lemma.
\begin{lemma}
    We can solve the \problem~in~$O(n)$ space and~$O(z+\log n + \mathrm{occ})$ time.
\end{lemma}

\section{Slice Tree Solution}\label{sec:slice_tree}

In this section, we show how to reduce the~$O(\log n)$ time overhead to $O(\log m)$.
This will originally give a space complexity of $O(n\log n)$. In the next section we show how to reduce the space to linear.
Recall that the additional~$O(\log n)$ time originates from the binary search in the case where after matching~$k-1$ phrases we arrive at a node, and the~$k^{th}$ phrase does not match any of the outgoing edges. In any other case, the solution from the previous section gives~$O(z+\mathrm{occ})$ time complexity. We use the solution from the previous section as a basis and show how to speed up the last step of matching the~$k^{th}$ phrase. 

We note that similar 
results follow from using \emph{$z$-fast tries} \cite{BelazzouguiBV10}, however, we present a direct and simple solution.

We use Karp-Rabin fingerprints and the ART tree decomposition, which we define next. 
\paragraph{Karp-Rabin Fingerprints}

For a prime~$p$ and an~$x\leq p$, the \emph{Karp-Rabin fingerprint}~\cite{karp1987efficient} of a substring~$\substring{S}{i}{j}$ is defined as 
\begin{align*}
    \phi_{p,x}(\substring{S}{i}{j})=\sum_{k=i}^{j-1} S[k]x^{k-i}\mod p.
\end{align*}
 Clearly, we have that if $\substring{S}{i}{j} = \substring{S'}{i'}{j'}$, then $\phi_{p,x}(\substring{S}{i}{j})=\phi_{p,x}(\substring{S'}{i'}{j'})$.  Furthermore, the Karp-Rabin fingerprint has the property that for any three strings $S$, $S'$ and $S''$ where $S=S'S''$, given the fingerprint of any two of those strings and constant additional information, the third one can be computed in constant time:
 \begin{lemma}\label{lem:fingerprints}
    Let $S$, $S'$ and $S''$ be three strings satisfying $S=S'S''$. Given the Karp-Rabin fingerprints $\phi$ of any two of the strings $S$, $S'$ and $S''$ we can calculate the third one as follows:
    \begin{align*}
        \phi(S)&= \phi(S')+x^{|S'|}\cdot \phi(S'')\mod p,\\
        \phi(S')&= \phi(S)-\frac{x^{|S|}}{x^{|S''|}}\cdot \phi(S'')\mod p,\\
        \phi(S'')&= \frac{\phi(S)-\phi(S')}{x^{|S'|}}\mod p.
    \end{align*}
 \end{lemma}
 It follows that given the fingerprints of all suffixes of a string $S$ as well as the exponents $x^j\mod p$ for all $j=0,\dots,n-1$, the fingerprint of any substring of $S$ can be computed in constant time.
 
 We assume that $p$ and $x$ are chosen in such a way that $\phi_{p,x}$ is \emph{collision-free} on substrings of $S$, that is, two distinct substrings of $S$ have different fingerprints. For details on how to construct $\phi_{p,x}$ see the paragraph on preprocessing. 
 We will from now on use the notation $\phi=\phi_{p,x}$.

\paragraph{ART Decomposition}
The ART decomposition of a tree by Alstrup et al.~\cite{alstrup1998marked} partitions a tree into a \emph{top tree} and several \emph{bottom trees} 
with respect to a parameter $\chi$: Every vertex~$v$ of minimal depth with no more than~$\chi$ leaves below it is the root of a bottom tree which consists of~$v$ and all its descendants. The top tree consists of all vertices that are not in any bottom tree. The following lemma gives a key property of ART trees:
\begin{lemma}[Alstrup et al.\cite{alstrup1998marked}]\label{lemmART}
The ART decomposition with parameter~$\chi$ for a rooted tree~$T$ with~$n$ leaves produces a top tree with at most~$\frac{n}{\chi + 1}$ leaves.
\end{lemma}

\subsection{The Slice Tree Decomposition}
The data structure we define in this section is only used for speeding up the matching of the $k^{th}$ phrase, assuming $k$ is the maximum number such that we matched $k-1$ phrases in the phrase trie. The overall idea is to use fingerprints to do an exponential search for the locus of $P$; since the locus is of depth at most $m$, this way, we will achieve a search time of $O(\log m)$. In order to do this, we will store fingerprints of substrings of $S$ for lengths of powers of two. For the detailed search of the remainder we divide the suffix tree into smaller trees, the \emph{slice trees}, where the heights are powers of two and increase with the depth in the tree. In order not to use too much space, we store an ART-decomposition of the slice trees. This way we can afford to store fingerprints for every string depth in the top tree and binary search within the bottom trees. This will result in an $O(n \log n)$ space data structure. However, in the next section we show how to reduce this space to linear.

In more detail, we store the space efficient phrase trie from the previous section for matching full phrases of the pattern. Additionally, we store the Karp-Rabin fingerprints for each suffix of~$S$, as well as the following \emph{slice tree decomposition} of the suffix tree of~$S$:
\begin{itemize}
    \item We store the suffix tree together with extra nodes at any position in the suffix tree that corresponds to a string depth that is a power of two. For each node we store the range in the suffix array of the leaves below.
    \item For each level of string depth~$2^i$, where~$i=0,\dots, \lfloor\log n \rfloor$, we store a static hash table with Karp-Rabin fingerprints of the substring in~$S$ from the root to every node of string depth~$2^i$.  As in section \ref{sec:compact_phrase_trie}, we use perfect hashing for all hash tables in this solution.
    \item For each node~$v$ at string depth~$2^i$ we define a \emph{slice tree} of order~$i$. The slice tree is the subtree rooted at~$v$, cut off at string depth~$2^i$, such that the string height of the slice tree is (at most)~$2^i$. 
    \item We compute an ART decomposition of each slice tree of order~$i$ with the parameter~$\chi$ set to~$\chi=2^{i}$. For each~$1\leq d<2^i$, we store a hash table with fingerprints corresponding to the substrings of length~$d$ starting at the root of the slice tree and ending in the top tree. Additionally, for every edge connecting a top tree node to a bottom tree root save the corresponding first letter in the suffix tree. 
\end{itemize}

\begin{figure}
    \centering
    \includegraphics[width=0.5\textwidth]{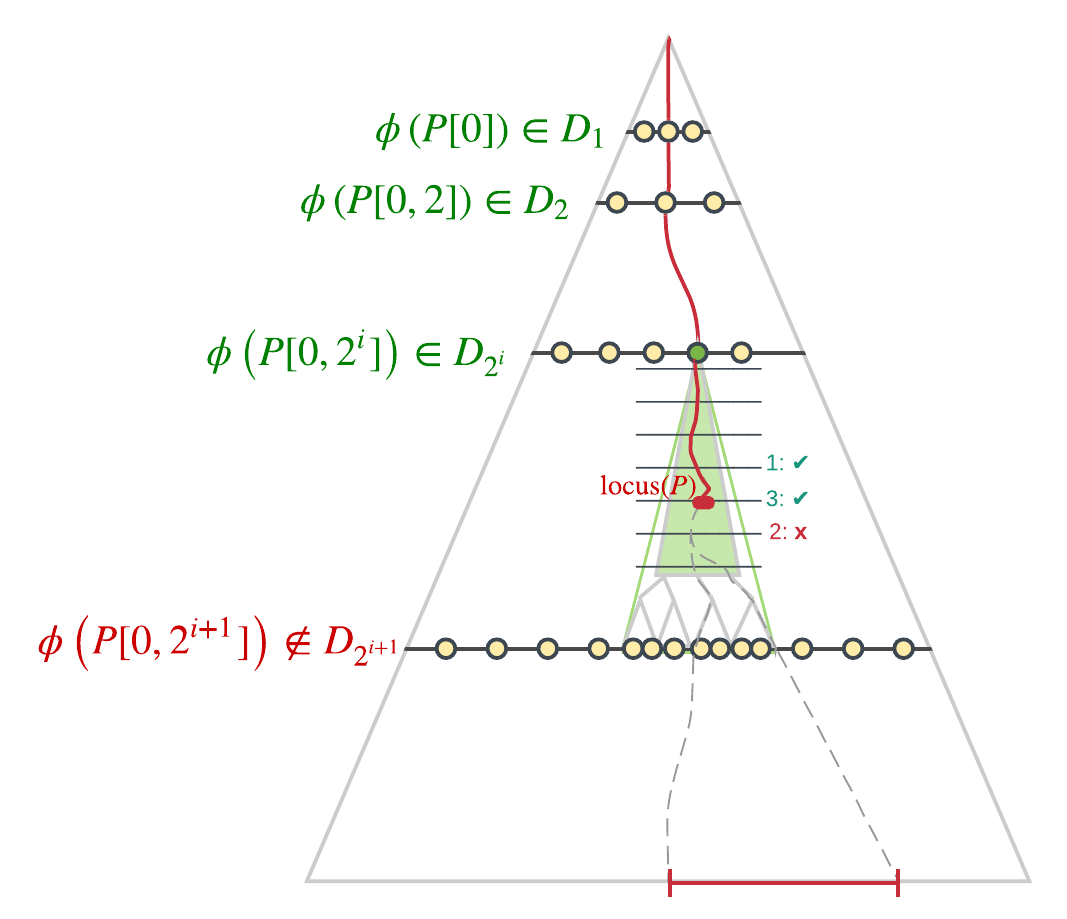}
    \caption{Matching in the slice tree: First, we find the lowest $i$ such that the fingerprint of a prefix of $P$ is present at level $2^i$. Then we go to the corresponding slice tree and binary search for fingerprints within the top tree. }
    \label{fig:my_label}
\end{figure}

\subsection{Algorithm}
To match $P$, we first match the full phrases in the phrase trie until we find the first phrase~$f_k$ which does not match any of the next phrases in the trie. If~$f_k$ is just a letter, as before, we are done. Otherwise~$f_k$ is represented by~$(r_k,l_k)$. For now, we assume $l_k\leq r_k$. At the end of the section we explain how to deal with self-referencing phrases.
Now:
\begin{itemize}
    \item We find the fingerprint~$\phi(\substring{P}{0}{p_k})=\phi(\substring{S}{i_0}{i_0+p_k})$, where $i_0$ is a leaf below the locus of $\LZ{P}$ in the phrase trie.
      Note that since $\substring{S}{i_0}{i_0+p_k}$ is a substring of $S$ and we stored the fingerprints of all suffixes of $S$ we can find its fingerprint in constant time via the fingerprints of the suffixes $S_{i_0}$ and $S_{i_0+p_k}$.
    \item In order to find the slice tree where the match ends, we do a linear search for the
    deepest matching fingerprint in the hash tables at the power of 2 levels in the following way: 
    \begin{itemize}
    \item For $j\in\left\{2^{\lceil\log p_k\rceil}-p_k,2^{\lceil\log p_k\rceil+1}-p_k,\dots,2^{\lfloor \log n \rfloor}-p_k\right\}$ and while $j<l_k$, we find the fingerprint of the prefix $\substring{f_k}{0}{j}=\substring{S}{i_0+p_k-r_k}{i_0+p_k-r_k+j}$ and look for $\phi(\substring{P}{0}{p_k})+x^{p_k}\phi(\substring{f_k}{0}{j})\mod p$ in the hash table of depth~$p_k+j$. 
    If $\phi(\substring{P}{0}{p_k})+x^{p_k}\phi(\substring{f_k}{0}{j})\mod p=\phi(S_i)$ for some~$i$, we check if $\phi(\substring{S}{i}{i+p_k})=\phi(\substring{P}{0}{p_k})$ to avoid false positives. We keep doing this until the first level where it is not present or the check fails.
    \item For the last level where there is a match, we find the corresponding node and the slice tree rooted at that node.  

    Note that this slice tree can be of order at most $\log m$.
    \end{itemize}
    \item Similarly as the linear search above, we now do an binary search for fingerprints on the levels in the top tree of the slice tree. For the lowest level in which there is a match in the top tree, find the corresponding position $v$. If this is an internal node without any off-hanging bottom trees or on an edge in the top tree then $\loc(P)= v$. Once we have found $\loc(P)$ we can easily find and return the occurrences as before. Otherwise, we check if the next letter in~$P$ matches any of the off-hanging bottom trees. We can find this letter in constant time by looking up its source in $S$. If it matches, we do a binary search for the longest match with the leaves of the bottom tree, which proceeds exactly as in the phrase trie solution, but restricted to any representative leaf below each bottom tree leaf. For each bottom tree leaf that has a longest match with $P$ report all suffix tree leaves below.
    
    \end{itemize}
    We note that the search algorithm is similar to prefix search in a z-fast trie, however, there are subtle differences. Let the \emph{2-fattest number} of an interval $[l,r]$ be the unique number of the form $b2^i$ in the interval such that $b$ is an integer and $i$ is maximal. If $l=0$, this is the largest power of two which is at most $r$. The search in a z-fast trie begins with computing the 2-fattest number in $[0,m-1]$, finds the first node in the trie that shares a prefix of that length with $P$ (if it exists), and then continues similarly to a binary search. The interval lengths in that search depend on the depths of the nodes and might not be powers of two, which is why the search relies on finding 2-fattest numbers. In contrast, our search can be seen as an exponential search for the length $p$ of the longest prefix of $P$ in the trie. That is, first, we find in $O(\log p)=O(\log m)$ steps the 2-fattest number in $[0,p]$; call this $q$. Then we binary search for $p$ in $[q,2q]$, where the interval lengths are powers of two.

\subsection{Correctness}
The correctness of matching the first $k-1$ phrases follows from the previous section.
Given that $k$ is the first phrase that does not match any of the next phrases in the suffixes, we argue for the search in the power of two levels in the suffix tree. We know that if~$\substring{P}{0}{p_k+j}=\substring{S}{i}{i+p_k + j }$ for some~$i$, then~$\phi(\substring{P}{0}{p_k})+x^{p_k}\phi(\substring{f_k}{0}{j})\mod p$ will be present in the hash table of level~$p_k +j$. Further, we chose~$\phi$ such that it has no false positives on substrings of~$S$, and (since we assume $l_k\leq r_k$) we know that both $\substring{P}{0}{p_k}$ and $f_k$ are substrings of $S$. Thus, by checking~$\phi(\substring{P}{0}{p_k})=\phi(\substring{S}{i}{i+p_k})$ separately, Lemma \ref{lem:fingerprints} implies that~$\substring{P}{0}{p_k+j}$ and~$\substring{S}{i}{i+p_k+j}$ are actually identical. Together, this means that by finding the biggest~$j$ such that~$p_k +j$ is a power of two and both conditions are fulfilled, we will find the slice tree that contains the end of the longest match.

Next, we argue for the detailed search within the slice tree. The argument for the binary search in the top tree is the same as for the search on the power of two levels. When we end the binary search, we found the position in the top tree of maximum depth that corresponds to a substring of~$S$ matching a prefix of~$P$. The longest match either ends there or in a bottom tree that is connected to this position. If there is more than one such bottom tree, the first letter on each edge will uniquely identify the bottom tree that contains the leaf or leaves with the longest match. If the longest match ends in a bottom tree, it is enough to do the binary search with any representative leaf in the suffix tree per leaf in the bottom tree,
since for any such leaf the prefix of a given length that ends within the bottom tree is the same.

\subsection{Analysis}
We use linear space for the phrase trie representation of the previous section and the fingerprints of the suffixes of~$S$.
Additionally, we use~$O(n\log n)$ space for the extra nodes and hash tables at the power of two levels. 

For each slice tree~$T$ of order~$i$ denote~$|T|$ the number of nodes in the slice tree and let $h=2^i$ be the maximal height of the slice tree. By Lemma~\ref{lemmART}, the top tree has at most~$|T|/h$ leaves. By the definition of the slice tree, each root-to-leaf path has at most~$h$ positions. As such, the hash tables for the top tree take up~$O(|T|)$ space. Furthermore we use constant space per leaf in the bottom tree. Each bottom tree leaf is a node in the suffix tree or an extra node, and each such node is a leaf in at most one bottom tree. So the total space for all slice trees is~$\sum_{T\textnormal{ is slice tree}}O\left(|T|\right)=O(\#\texttt{nodes in suffix tree }+\texttt{ extra nodes})= O(n\log n)$.

For the time complexity, as before, we use~$O(z)$ time for matching in the phrase trie. Since we stored the fingerprints of all suffixes of~$S$, the fingerprint of any substring of~$S$ can be found in constant time. 

For the linear search of fingerprints in the suffix tree, note that the last phrase of~$P$ is at most~$m$ long. This means we stop the search after checking at most~$\log m$ power of 2 levels, and a check can be done in constant time.

After the linear search we end up in a slice tree of order at most~$\log m$, which means~$h\leq m$. It follows that the binary search in the top tree uses time at most~$O(\log h)= O(\log m)$. Further, by the definition of the ART decomposition, every bottom tree has no more than~$h\leq m$ leaves, and as such the binary search in the bottom tree uses no more than~$O(\log m)$ operations.  

In total, this gives us a time complexity of~$O(z+\log m + \mathrm{occ})$.

        \subsection{Handling Self-Referencing Phrases}
Now we describe how to use the slice tree for matching phrase $f_k$ in the case that $f_k$ is self-referencing. Assume we already matched the first $k-1$ phrases in the phrase trie. We will show how to match $f_k$ in the slice tree in three steps: First, we will show how to construct the fingerprints of all prefixes of $f_k$ of length a power of two  in  $O(\log m)$ time. Then we show how this information enables us to find the longest match in the slice tree in $O(\log m)$ time. Finally, we will show how to check for false positives. 
       
Let $f_k=(r_k,l_k)$ with $l_k>r_k$ and let $p_k$ denote the starting position of $f_k$ in $P$. Since we have matched $P$ up to position $p_k$, the first $r_k$ letters of $f_k$ are given by a substring of $S[j+p_k-r_k,j+p_k]$, where $j$ is any leaf below the last position we matched in the phrase trie. Call this substring $\Per$. Since $f_k$ is self-referencing, it is periodic, that is, it is a concatenation of copies of $\Per$ (where the last one might be incomplete). 
        
\begin{figure}
    \centering
    \includegraphics[width=0.55\textwidth]{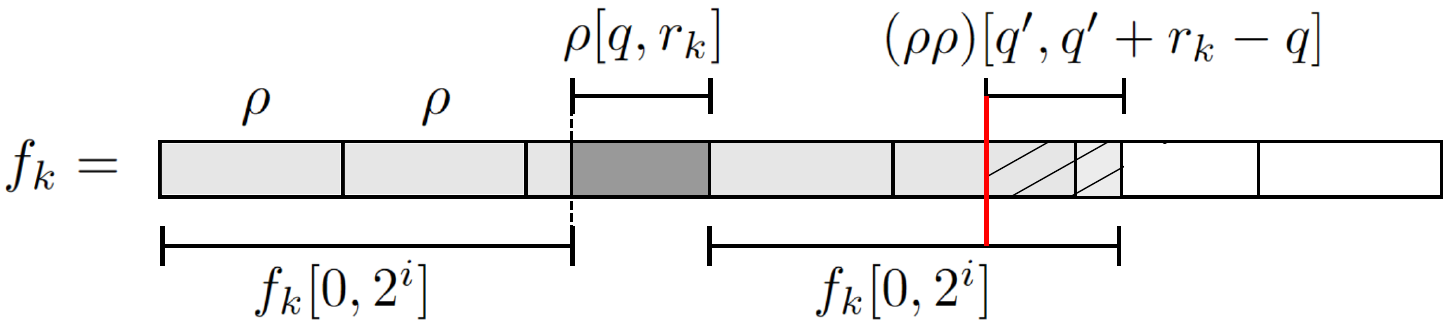}
    \caption{The prefix of length $2^{i+1}$ can be constructed from the prefix of length $2^i$ and substrings of $\rho$ resp.~$\rho\rho$.}
    \label{fig:selfreferencing}
\end{figure}

        \paragraph{Finding all fingerprints of power of two prefixes.}
        First, we find and store the fingerprints of all prefixes of $f_k$ where the length is a power of two, by repeatedly doubling and using Lemma \ref{lem:fingerprints}:
        
        While $2^i\leq r_k$, $f_k[0,2^i]$ is a substring of $S[j+p_k-r_k,j+p_k]$ and we can find its fingerprint in constant time.
        Then, given the fingerprint of $f_k[0,2^{i}]$, while $2^{i+1}\leq l_k$, we can find the fingerprint of $f_k[0,2^{i+1}]$ in constant time: 
        Note that $f_k[0,2^{i+1}]$ is a concatenation of $f_k[0,2^{i}]$, some suffix of $\Per$ to ``fill up" until the end of the next period, and a prefix of $f_k[0,2^{i}]$ (see Figure~\ref{fig:selfreferencing}). This last prefix can be constructed from $f_k[0,2^{i}]$ by subtracting a substring of $\Per\Per$. Thus the fingerprints of $f_k[0,2^{i+1}]$ can be computed by combining the fingerprints of these substrings.   
        More precisely, let $q=2^i \mod r_k$. That means, at the end of $f_k[0,2^{i}]$ there is a period cut off after $q$ characters. So if we concatenate $f_k[0,2^i]$ and $\Per [q,r_k]$, we get a prefix of $f_k$ that consists of full periods only. If we then append $f_k[0,2^i]$, we get a periodic string of length $2^{i+1}+r_k-q$, so we need to ``cut off" the last $r_k-q$ elements. Let $q'=2^{i+1}\mod r_k$. The substring we cut off corresponds to $(\Per\Per) [q',q'+r_k-q]$. 
        
By assumption, we know the fingerprint of $f_k[0,2^i]$. We can find the fingerprints of substrings of $\Per$ in constant time by translation to $S$, and substrings of $\Per\Per$ as a concatenation of at most two substrings of $\Per$. Using Lemma \ref{lem:fingerprints}, we can thus find the fingerprint of $f_k[0,2^{i+1}]$ in constant time. The total time for finding the fingerprints of all prefixes of $f_k$ of length a power of two thus takes $O(\log m)$ time.
        \paragraph{Matching in the Slice Tree.} Once we have stored all fingerprints of prefixes of $f_k$ of length a power of two, we can find the fingerprint of \emph{any} substring of $f_k$ that is a power of two \emph{in constant time}. Let $s = f_k[a,2^i+a]$ be such a substring. There are two cases:
        \begin{enumerate}
        \item If the substring starts at a position $j\cdot r_k$ for some $j$, then due to the periodicity $f_k[a,2^i+a] = f_k[0,2^i]$, and thus it has the same fingerprint as the prefix of the same length.
        \item Otherwise, due to the periodicity of $f_k$ the substring is equal to some $f_k[b,2^i+b]$ for $b<r_k$. Now $f_k[b,2^i+b]$ can be constructed by concatenating the suffix $f_k[b,2^i]$ of $f_k[0,2^i]$ with a substring of $\Per\Per$, and thus we can compute its fingerprint in constant time as described above. 
        \end{enumerate}
 Now, we can match in the slice tree in the following way:
        We find the fingerprint of the prefix of $f_k$ from its starting position in the suffix tree to the next power of two level. Since this is a concatenation of at most $\log m$ substrings of length which are a power of two, we can do this in $O(\log m)$ time.
        Then, we can ``jump" between power of two levels in additional constant time. That is, we only need constant time for each step in the exponential search over the power of two levels, because we only need to add or subtract the fingerprint of a substring that is a power of two.
        Similarly, when binary searching within a slice tree, we always make steps of length that are a power of two, hence, every step can be done in constant time. 
        Thus, just as in the case for non self-referencing phrases, the fingerprint search takes a total of $O(\log m)$ time. 
        \paragraph{Checking for false positives.} Having found the longest match of fingerprints within the slice tree, we can check for false positives in $O(\log m)$ time, by a similar repeated doubling trick as before. Let $j$ be a leaf below the last matched position in the slice tree.
        We will iteratively check if prefixes of power of two lengths of $f_k$ are actually substrings of $S$ and match the corresponding positions in $S_j$. At every step, we check if the partial fingerprints match the fingerprints of corresponding substrings of $S_j$, and use that $\phi$ is constructed such that distinct substrings of $S$ have different fingerprints.
        
 In detail, let $i=0,\dots, \lfloor\log l_k \rfloor$. While $2^i\leq r_k$, we know that $f_k[0,2^i]$ is a substring of $S$. We check if $\phi(f_k[0,2^i])=\phi(S_j[p_k,p_k+2^i))$. If yes, then the substrings are the same.
For $2^i>r_k$, assuming we have verified that $f_k[0,2^i]$ is a substring of $S$ and is equal to $S_j[p_k,p_k+2^i]$, we can check if $f_k[0,2^{i+1}]$ is equal to $S_j[p_k,p_k+2^{i+1}]$ in the following way: 
Since $f_k[0,2^{i+1}]= f_k[0,2^{i}]\Per [q,r_k]f_k[0,2^i-(r_k-q)]$, we know that all substrings on the right are substrings of $S$. Additionally, we know that $f_k[0,2^{i}]=S_j[p_k,p_k+2^i]$. We check if $\phi(\Per [q,r_k])=\phi(S_j[p_k+2^i, p_k+2^i+r_k-q])$ and if $\phi(f_k[0,2^i-(r_k-q)])=\phi(S_j[p_k+2^i+r_k-q, 2^{i+1}])$. We can compute all these fingerprints in constant time, and since we always compare fingerprints of substrings of $S$, we know that if the fingerprints are the same then the substrings are the same. Hence, in that case, $f_k[0,2^{i+1}]$ is equal to $S_j[p_k,p_k+2^{i+1}]$, which also means it is a substring of $S$. 
After we have verified the prefix $f_k[0,2^{\lfloor \log l_k\rfloor}]$, the full $f_k$ is again a concatenation of $O(\log m)$ strings we have already verfied to be substrings of $S$, and we can check them in the same way.

We arrive at the following result:

\begin{lemma}\label{lemma_slice_tree}
The slice tree solution solves the \problem\ in~$O(n \log n)$ space and~$O(z+\log m + \mathrm{occ})$ time.
\end{lemma}

\subsection{Preprocessing}
We now describe how to construct the data structure, especially, how to choose the fingerprint function $\phi_{p,x}$.
\paragraph{Choosing $\phi_{p,x}$}
In \cite{karp1987efficient} it is shown that for good choice of $p$ and uniformly random $x\in \mathbb{Z}_p$, the fingerprint function $\phi_{p,x}$ is collision-free on substrings of $S$ with high probability. 

 For any choice of $x$, we can check if $\phi_{p,x}$ is collision-free in expected time $O(n^2)$ and $O(n)$ additional space. Since in the algorithm, we only ever compare the fingerprints of substrings that have the same length, it is enough to make sure $\phi_{p,x}$ is collision free on substrings of a given length $l$ (we can also trivially extend such a fingerprint to a fingerprint function that is collision free on all substrings of $S$, but we don't need to). Now, for every $1\leq l\leq n-l$, we simply compute all $\phi_{p,x}(S[i,i+l])$ for $0\leq i \leq n-1$ and keep a dictionary using universal hashing \cite{CarterW77}, to check if any two substrings of length $l$ map to  the same fingerprint. Then we discard the dictionary. 

 Since $\phi_{p,x}$ is collision-free with high probability, we will find a collision-free $\phi_{p,x}$ in exptected constant number of tries, which gives an expected $O(n^2)$ running time for finding a collision-free $\phi_{p,x}$.
\paragraph{Faster Construction for \LZss\ without self-references} If we use \LZss\ without self-referencing, we can match $f_k$ in the phrase trie and then check for false positives in $\log m$ time using only fingerprints which are a power of two long, in the following way: We can divide $f_k$ into $\log m$ substrings of lengths which are a power of two. Since $f_k$ is not self-referencing, all these are substrings of $S$. We compare their fingerprints with the fingerprints of the corresponding substrings in the potential match. Thus, in the case of non-self referencing, it is enough that $\phi_{p,x}$ is collision free on substrings of $S$ that have a length that is a power of two, and by the same strategy as before, such a $\phi_{p,x}$ can be constructed in $O(n\log n)$ expected time (see also Bille et al. \cite{bille2015longest}).
\paragraph{Final Construction} Once we have chosen $\phi_{p,x}$, we precompute the fingerprints of all suffixes of $S$, which can be done in $O(n)$  time: First, we compute $x^i\mod p$ for all $0\leq i <n$ in $O(n)$ time, then we use Lemma~\ref{lem:fingerprints} to compute $S_{n-1}, S_{n-2},\dots, S_0=S$, in that order.  After we have stored the fingerprints of all suffixes of $S$, computing the fingerprint of any substring can be done in constant time. Using perfect hashing~\cite{FKS1984}, we can build all dictionaries in expected time linear to their size, in total, $O(n\log n)$ expected time. The ART decompositions can be constructed in time linear in the nodes, i.e. $O(n\log n)$ total worst case time. Together with the preprocessing time from Sections \ref{sec:phrase_trie} and \ref{sec:compact_phrase_trie}, we can construct the full data structure in
\begin{enumerate}
    \item $O(n^2+\sum_{i=0}^{n-1} z_i\log \log n)$ expected time if we allow self referencing;
    \item $O(n\log n+\sum_{i=0}^{n-1} z_i\log \log n)$ expected time if we do not allow self referencing.
\end{enumerate}

\section{Saving Space}\label{sec:saving_space}

For the solution above, we constructed~$O(n\log n)$ slice trees. By the way we defined them, note that any internal node in a slice tree has to be an original node from the suffix tree. Since there are only~$O(n)$ such nodes, we conclude that many of the slice trees consist of a single edge. We will show that by removing those, we can define a linear space solution that gives the same time complexity as in Lemma~\ref{lemma_slice_tree}.

\subsection{The Data Structure}
We start with the slice tree solution. Call every suffix tree edge that contains two or more extra nodes a \emph{long edge}. For every long edge, delete every extra node except the first and last, which we call~$\firstnode$ and~$\lastnode$. For every deleted node also delete the additional information stored for their slice trees, and their corresponding entries in the power of two hash tables. For each long edge, store at the hash table position of~$\firstnode$ additionally the information that it is on a long edge, how long that edge is, and a leaf below it.

\subsection{Algorithm}
The algorithm proceeds almost the same way as before. The only change is that in the linear search of power of two levels, when we match with a node that is $\firstnode$ of a long edge, jump directly to the last power of two level that is before the end of the edge. If the fingerprint is present, proceed normally, otherwise, the longest match ends on that edge and we do a single \LCP~query between the source of the phrase in $S$ and the stored leaf to find its length.

\subsection{Correctness} If we do not encounter any long edges, nothing changes. If a long edge is entirely contained in the match, we will first find~$\firstnode$ and then jump directly to the last power of two level on that edge, where we will find~$\lastnode$, and then continue as before. If the longest match ends on a long edge, there are two cases:
\begin{enumerate}
    \item The longest match ends before~$\firstnode$ or after~$\lastnode$: this means that by doing the linear search we find the slice tree that the longest match ends in, thus everything follows as before.
    \item The longest match ends between~$\firstnode$ and~$\lastnode$: In this case, we will find a matching fingerprint at the level corresponding to~$\firstnode$ but no matching fingerprint at the level corresponding to~$\lastnode$, which means we will use \LCP~to find the longest match with a leaf below~$\firstnode$. Since the match ends on that edge, this gives us the correct length and position. 
\end{enumerate}
\subsection{Analysis} For space complexity, note that we only keep original nodes from the suffix tree, plus at most two extra nodes per edge, so a linear number of nodes in total. Since the space used for the slice trees and power of two hash tables is linear in the number of nodes, the total space consumption is linear. The time complexity does not change. This concludes the proof of Theorem \ref{thm:mainresult}.

\section{LZ78-compressed patterns}\label{sec:lz78}
 As an extension to our result, we show that a very similar solution solves the problem for the Lempel-Ziv 1978 (\LZse) compression scheme \cite{ZivL78}.

    \paragraph{LZ78}\label{subsec:lz78}
Given an input string $S$ of length $n$, the \LZse\ parsing divides $S$ into $z$ substrings $f_1, f_2,\ldots,f_z$, called phrases, in a greedy left-to-right order. The $i^{th}$ phrase $f_i$, starting at position $p_i$ is either (a) the first occurrence of a character in $S$ or (b) the longest substring that is equal to a phrase $f_j$, $j<i$, plus the next character.
Note that this choice is unique.
To compress $S$, we can then replace each phrase $f_i$ of type (b) with a pair $(j,\alpha)$ such that $j$ is the index of the phrase $f_j$, and $\alpha$ is the next character.

\paragraph{Data structure} The phrase trie with respect to \LZse\ is defined completely analogously to Section \ref{sec:phrase_trie}; the only difference is that the suffixes of $S$ are \LZse\ compressed. The representation from Section \ref{sec:compact_phrase_trie} can be applied directly. The final data structure consists of the efficient representation of the \LZse\ phrase trie together with the (unchanged) slice tree solution defined in Sections \ref{sec:slice_tree} and \ref{sec:saving_space}.
\paragraph{Algorithm} 
    When matching in the phrase trie, we build up a dictionary mapping LZ78 phrases to substrings in $S$. That is, assume $p_k$ is the starting position of $f_k$ in $P$, and we fully matched up until the end of $f_k$ in the phrase trie. Then, we add an entry to the dictionary where the key is the phrase index $k$ and the value is a pair $(j+p_k,j+p_k+|f_k|)$, where $j$ is a leaf below the current position in the phrase trie.

 Assume we have matched up to a position $p_k$ of $P$. Let $f_{k}=(f_{a},\alpha)$ be a new phrase in $P$ and $f'_{k}=(f'_{b},\beta)$ be a new phrase in the phrase trie, and let $(i,l)$ be the dictionary entry at $a$. That is, $i$ is a starting position of $f_a$, and $l$ is the length of $f_a$. Further, let $j$ be a leaf below the current poisition in the phrase trie. Then, similarly as in Lemma \ref{lcemma}, we have:
    \begin{align} \LCP(P,S[j,n])&\geq p_k+\min(\LCP(S[i,n],S[j+p_k,n]),l)\label{form:lz78-match}\\
    \LCP(P,S[j,n])&= p_k+\min(\LCP(S[i,n],S[j+p_k,n]),l), \textnormal{~~if~}f_{a}\neq f'_{b}.\label{form:lz78-match2}\end{align}
 To see that (\ref{form:lz78-match}) is true note that by definition of $i$ and $l$, $f_a=S(i,i+l)$.
If $f_a\neq f'_b$, then (\ref{form:lz78-match2}) holds by the greedy parsing.

 Thus, the main property we need for matching within the (blind) phrase trie is preserved. Note that unlike the version of \LZss\ we use in this paper, an LZ78 phrase always includes an extra letter at the end. However, that is not an issue, since we can always access the next character in the phrase trie in constant time through $S$. For the slice tree solution, we only need that the last phrase is encoded as a substring of $S$, which is given by the dictionary. Thus, all results from the previous sections generalize to \LZse. 
    
    We arrive at the following result:
    \begin{theorem}
\label{thm:notmainresult}
We can solve the \problem\ for \LZse-compressed patterns in~$O(n)$ space and~$O(z+\log m + \mathrm{occ})$ time, where $n$ is the length of the indexing string, $m$ is the length of the pattern, and $z$ is the number of phrases in the \LZse\ compressed pattern.
\end{theorem}

\section{Open Problems}
We have introduced the \problem\ and provided a solution achieving almost optimal bounds for \LZss\ compressed patterns. Further, we have shown that the results extend to the \LZse\ compression scheme. At the same time, these results open some interesting directions for further research: 
\begin{itemize}
    \item Our results are optimal for the \LZss\ variant without self-referencing. An interesting open question is if there is a way to get optimal time for the self-referencing variant, that is, get rid of the additional $O(\log m)$ time overhead.
    \item Similarly, it would be interesting to see if we can get rid of the $O(n^2)$ expected construction time for self-referencing while still giving a Las Vegas algorithm.
    \item It would be interesting to consider the \problem\ for other compression schemes. Especially, is there a way to compress multiple patterns that allows a similar tradeoff? 
    \item Finally, the related problem where the indexing string and the pattern are \emph{both} compressed is especially interesting for practical use cases, because in many practical scenarios, the indexing string will be very long.
\end{itemize}

\bibliographystyle{plainurl}
\bibliography{References}

\end{document}